\documentclass{article}

\PassOptionsToPackage{numbers, compress}{natbib}
\usepackage[final]{neurips_2021_ml4ps}

\usepackage[utf8]{inputenc} 
\usepackage[T1]{fontenc}    
\usepackage{hyperref}       
\usepackage{url}            
\usepackage{booktabs}       
\usepackage{amsfonts}       
\usepackage{nicefrac}       
\usepackage{microtype}      
\usepackage{xcolor}         
\usepackage{float}
\usepackage{graphicx}

\usepackage[ruled]{algorithm2e}
\SetArgSty{textrm}
\DontPrintSemicolon

\makeatletter
\newcommand{\algorithmfootnote}[2][\footnotesize]{%
  \let\old@algocf@finish\@algocf@finish
  \def\@algocf@finish{\old@algocf@finish
    \leavevmode\rlap{\begin{minipage}{\linewidth}
    #1#2
    \end{minipage}}%
  }%
}
\makeatother

\newcommand\pcr{\textit{Partial Contribution Representation}}
\newif\ifanon
\anonfalse

\newif\ifscratch
\scratchtrue
\usepackage{soul}

\newcommand{\anontext}[1]{\ifanon \textbf{[removed for anonymity]} \else #1 \fi}

\title{Partial-Attribution Instance Segmentation for Astronomical Source
       Detection and Deblending}

\author{%
  Ryan Hausen \\
  Department of Computer Science and Engineering\\
  University of California, Santa Cruz\\
  Santa Cruz, CA 95064 \\
  \texttt{rhausen@ucsc.edu} \\

  \And
  Brant Robertson \\
  Department of Astronomy and Astrophysics\\
  University of California, Santa Cruz\\
  Santa Cruz, CA 95064 \\
  \texttt{brant@ucsc.edu} \\
}

\begin{document}

\maketitle

\begin{abstract}
  Astronomical source deblending is the process of separating the contribution
  of individual stars or galaxies (sources) to an image comprised of multiple,
  possibly overlapping sources. Astronomical sources display a wide range of
  sizes and brightnesses and may show substantial overlap in images.
  Astronomical imaging data can further challenge off-the-shelf computer vision
  algorithms owing to its high dynamic range, low signal-to-noise ratio, and
  unconventional image format. These challenges make source deblending an open
  area of astronomical research, and in this work, we introduce a new approach
  called Partial-Attribution Instance Segmentation that enables source detection
  and deblending in a manner tractable for deep learning models. We provide a
  novel neural network implementation as a demonstration of the method.
\end{abstract}

\section{Introduction}
\label{sec:intro}

Astronomical images can contain tens of thousands of stars and galaxies
(sources). Forthcoming telescopes including the Vera Rubin Observatory
\citep{ivezic2008a,ivezic2019a}, James Webb Space Telescope
\citep{williams2018a}, and Nancy Grace Roman Space Telescope
\citep{spergel2013a,spergel2015a} will push the current limits of observational
astronomy and dramatically increase the number of sources to analyze. To measure
accurate properties for these sources, we must {\it detect} sources by
identifying statistically significant local maxima in an image and {\it deblend}
sources by isolating the potentially overlapping flux distributions of each
object.
Consider a background-subtracted astronomical image $\mathcal{I} \in
\mathbb{R}^{h \times w \times b}$ in which $n$ sources are
observed, where $h$ is the height, $w$ is the width, and $b$ indicates
the number of astronomical passbands.
The image $\mathcal{I}$ can be decomposed into a sum of
individual object contributions as
\begin{equation}
  \label{eq:image_sum}
  \mathcal{I} = \sum_{i=1}^N S_i + \epsilon
\end{equation}
\noindent
where $S_i \in \mathbb{R}^{h \times w \times b}$ represents the flux contributed
to $\mathcal{I}$ by source $i$, and $\epsilon \in \mathcal{N}(0, \sigma)$ is the
approximate noise distribution in the image.
The process of decomposing an image into the form of
Equation \ref{eq:image_sum} represents the core challenge of source deblending.
This submission presents a deep learning-based method
to perform detection and deblending on astronomical images.

\subsection{Related Work}
\label{sec:related-work}

Source detection and deblending are well-studied problems in astronomy, and many
approaches have been developed. Below, we highlight some popular and recent
methods for source detection and deblending and point the interested reader to
the review by \citet{masias2012}.

Detection and deblending methods can be characterized by their \textit{detection
capacity} and \textit{deblend type}. The detection capacity represents the
number of sources a method can detect within a single image. For Equation
\ref{eq:image_sum}, a detection capacity of $N$ would indicate that a method
could detect all sources appearing in an image. The deblend type indicates how
the flux in a single pixel may be split between overlapping (blended) sources. A
\textit{disjoint} deblender assigns all flux in a pixel to a single source
exclusively. An \textit{intersecting/discrete} deblender can assign the flux to
more than one source with uniform weighting across pixels. Finally, an
\textit{intersecting/continuous} deblender can assign the flux to more than one
source with variable weighting across pixels.

Astronomical analysis methods vary in their detection and deblending methods.
\citet{bertin1996} introduced SExtractor that uses a convolution and
thresholding approach for detection, and an isophotal analysis using binned
pixel intensity for deblending.
\citet{hausen2020} introduced Morpheus, a U-Net \cite{ronneberger2015} style
convolutional neural network (CNN) model that filters out background pixels,
uses a thresholding approach for detection, and combines watershed and peak
finding algorithms for deblending.
Another U-Net based model called \texttt{blend2mask} \citep{boucaud2020}
performs detection and deblending using the U-Net alone.
\citet{reiman2019} use a modified Super-Resolution GAN (SRGAN) \citep{ledig2016}
to deblend overlapping sources.
\citet{buke2019} trained a Mask R-CNN \citep{he2017} model to detect and deblend
sources.
SCARLET \citep{melchior2018} deblends sources using constrained matrix
factorization.

Table \ref{table:methods-summary} summarizes the features of these previous
methods, none of which have a detection capacity of $N$ and an
\textit{intersecting/continuous} deblend type. We now present a deep
learning-based \textit{intersecting/continuous} deblending algorithm with a
detection capacity of $N$.

\begin{table}
  \caption{Detection and deblending method categorization}
  \label{table:methods-summary}
  \centering
  \begin{tabular}{lcc}
    \toprule
    Name                                        & Detection Capacity & Deblend Type\\
    \midrule
    SExtractor\citep{bertin1996}                & $N$             & Disjoint                \\
    Morpheus\citep{hausen2020}                  & $N$             & Disjoint                \\
    Mask R-CNN\citep{buke2019}                  & $N$             & Intersecting/Discrete   \\
    \texttt{blend2mask2flux}\citep{boucaud2020} & $2$             & Intersecting/Discrete   \\
    Modified SRGAN\citep{reiman2019}            & 0               & Intersecting/Continuous \\
    SCARLET\citep{melchior2018}                 & 0               & Intersecting/Continuous \\
    This Work                                   & $N$             & Intersecting/Continuous \\
    \bottomrule
  \end{tabular}
\end{table}

\section{Partial-Attribution Instance Segmentation}
\label{sec:partial-instance-segmentation}

Partial-Attribution Instance Segmentation (PAIS) is a new extension of the instance
segmentation paradigm that allows for weighted, overlapping
segmentation maps. PAIS differs from other segmentation schemes
like cell segmentation \citep{zhou2020}, interacting surface segmentation
\citep{xie2020}, and amodal instance segmentation \citep{li2016}.
PAIS aims to isolate objects appearing in an image while preserving their
measurable quantities within areas of
overlap. For PAIS, we can approximate Equation \ref{eq:image_sum}
as

\begin{equation}
  \label{eqn:instance-seg}
  \tilde{\mathcal{I}} = \sum_{i=1}^N M_i \odot \mathcal{I}
\end{equation}
\noindent
where $\tilde{\mathcal{I}} \in \mathbb{R}^{h \times w \times b}$ estimates the
background-subtracted flux image $\mathcal{I}$ in Equation \ref{eq:image_sum},
$M_i \in [0,1]^{h \times w \times b} st. \sum_i^N M_{i,jkl}=1$ constitutes the
pixel-level fractional contribution of source $i$ to $\mathcal{I}$, and $\odot$
symbolizes the Hadamard product. Equation \ref{eqn:instance-seg} is tractable
for deep learning models, allowing the model to learn the bounded quantities
$M_{i}$ rather than the unbounded source images $S_{i}$. The $N$ number of
sources setting the upper limit of the sum in Equation \ref{eqn:instance-seg}
can differ for each image.

To construct a PAIS format that can be represented by a CNN, we have to
construct an encoding for the $M_i$ in Equation \ref{eqn:instance-seg}. Inspired
by \citet{cheng2019} and \citet{kendall2017}, we propose an encoding for the
$M_i$ components called \pcr{} (PCR). The goal of PCR is to encode, for any
single pixel $(j,k,l)$, the fractional contribution to its intensity from the
closest $n$ sources. Using PCR, a variable number $N$ of sources can be encoded
per image. PCR consists of three tensors: the Center-of-mass $C^c \in \{0,
1\}^{h \times w}$, Contribution-vectors $C^v \in \mathbb{R}^{h \times w \times n
\times 2}$ and Contribution-maps $C^m \in [0,1]^{h \times w \times b \times n}$.
The center-of-mass encodes the locations of all the sources in an image. For any
pixel, we set $C^c_{jk}=1$ if that location indicates the center of a source and
$C^c_{jk}=0$ otherwise. The contribution-vector $C^v_{jk}$ encodes the Cartesian
offset to the closest $n$ sources. The contribution-map $C^m_{jkl}$ connects the
fractional contribution of the $n$ sources with the associated
contribution-vectors $C^v_{jk}$. The fixed dimensionality of
$C^c$, $C^v$, and $C^m$ make PCR tractable for deep learning algorithms.

\section{Our Approach}
\label{sec:our-approach}

Our approach consists of making a PAIS dataset leveraging PCR and is implemented
using a novel neural network architecture. We summarize our dataset, model, and
training method below.

\subsection{Dataset}
\label{sec:dataset}

To generate the PAIS input samples, we used the Hubble Legacy Fields (HLF)
GOODS-South F160W ($1.6\mu\mathrm{m}$) flux images \citep{illingworth2016},
along with the 3D-HST source catalog \citep{momcheva2016}. The HLF images were
split into training and test sets of $256 \times 256$ pixel subregions, with
2,000 training samples and 500 test samples. The input labels, as described in
Section \ref{sec:partial-instance-segmentation}, consist of the center-of-mass
images $C^c$, the contribution-vectors $C^v$, and the contribution-maps $C^m$.
The center-of-mass training images are generated in a manner similar to
\citet{cheng2019}, by placing pixelated 2D Gaussians with standard deviation
$\sigma=8(\textrm{pixels})$ at the locations of sources in the 3D-HST catalog.
The contribution-vectors, an extension to the method by \citet{cheng2019}, are
generated from the Cartesian offset to the nearest $n=3$ sources to each pixel.
The contribution-maps require the $M_i$ values from Equation
\ref{eqn:instance-seg}. To determine $M_i$, we use SCARLET \citep{melchior2018}
with the F125W, F160W, F606W, and F850LP flux and weight images from the HLF
GOODS-South data and the TinyTim point-spread functions \citep{krist2011} to
deblend the sources from the 3D-HST catalog. We then use PCR to encode the $M_i$
from SCARLET. The complete dataset generation routine can be found in our
project repository
(\anontext{\url{https://github.com/ryanhausen/morpheus-deblend/}}).

To evaluate the efficacy of PCR to encode $M_i$, we define two metrics. We use
the mean difference between the total flux determined by the SCARLET encoded
$M_i$ for each input source and that recovered by our encoding. We also use a
two-sample Kolmogorov–Smirnov (KS) test to compare the normalized cumulative
surface brightness profile within the radius encompassing 90\% of the total flux
of each source to evaluate the encoding of the spatial flux distribution. Table
\ref{table:pcr-efficacy} reports the results and demonstrates that PCR encoding
approximately preserves both the total flux and the spatial flux distribution
for each source. With this verification, we can train a network to recover the
PCR encoding for each input HST F160W image.

\begin{table}
  \caption{\pcr{} encoding efficacy}
  \label{table:pcr-efficacy}
  \centering
  \begin{tabular}{ll}
    \toprule
    Test & Value \\
    \midrule
    Total Source Flux [e/s] (MAE) & $1.97 \pm 15.43$  \\
    Two-Sample KS Test p-value          & $0.93\pm0.22$     \\
    \bottomrule
  \end{tabular}
\end{table}

\subsection{Model}
\label{sec:model}

\begin{figure}
  \centering
  \includegraphics[width=\textwidth]{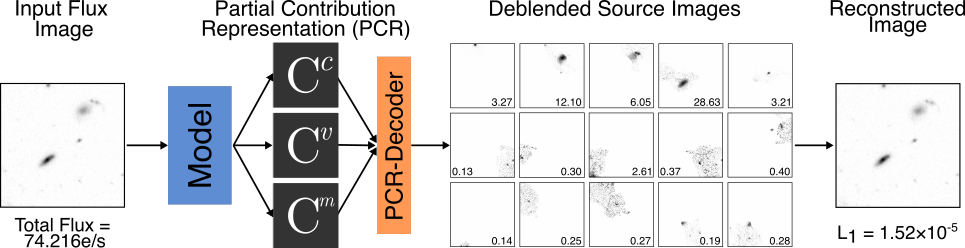}
  \caption{End-to-end example using our method to detect and deblend sources.
           Starting from the left: A flux image is input to the Model (see
           Section \ref{sec:model}). The Model outputs the deblended image in
           the \pcr{} (PCR; see Section
           \ref{sec:partial-instance-segmentation}).The output from the model is
           then decoded using the non-learned PCR Decoder algorithm into
           separate deblended source images. The deblended source images have
           their total flux within $r_{90}$ annotated. The deblended source
           images are then added together to generate the reconstructed image
           which has an $L_1$ total flux difference of $1.52 \times 10^{-5}$
           with the original input image.}
  \label{fig:example}
\end{figure}

To recover the PCR for training images, we developed a novel neural network
architecture inspired by \citet{cheng2019}, based on the Fast Attention Network
\citep{hu2020a} and implemented in TensorFlow \citep{tensorflow2015}. The model
features two decoders that share a single encoder. The first decoder, called the
\textit{spatial decoder}, predicts values for $C^c$ and $C^v$. The second
decoder, called the \textit{attribution decoder}, predicts values for $C^m$. The
complete model code can be found in the repository for this project
(\anontext{\url{https://github.com/ryanhausen/morpheus-deblend/}}). An
end-to-end example of the model can be seen in Figure \ref{fig:example}.

\subsection{Training}
\label{sec:training}

To train the model to recover the PCR of the input images, we use the Adam
Optimizer \citep{kingma2014} with a learning rate of $5 \times 10^{-5}$,
$\beta_1=0.9$, $\beta_2=0.9999$, $\epsilon=1\times 10^{-7}$, and a batch size of
100. The model was trained for 1000 epochs using an NVIDIA V100 32GB GPU, taking
31 hours. The loss function for the model is composed of three functions. The
\textit{spatial decoder} outputs for $C^c$ and $C^v$ are penalized according to
mean squared error (MSE) and the mean absolute error (MAE), respectively. The
\textit{attribution decoder} output $C^m$ is penalized using cross-entropy loss
with an additional entropy regularization term. In practice, we found that the
additional entropy regularization helped incentivize the network to learn
information about multiple sources in $C^m$. Each loss term is weighted and
combined into a single loss function described by

\begin{equation}
  \mathcal{L}_{\mathrm{total}} = \lambda_{C^c}\mathcal{L}_{C^c}
                                 + \lambda_{C^v}\mathcal{L}_{C^v}
                                 + \lambda_{C^m}\mathcal{L}_{C^m}
                                 + \lambda_{S}\mathcal{L}_{S},
\end{equation}
\noindent
where $\mathcal{L}_{C^c}$ is the MSE loss calculated between the model output
and input label with $\lambda_{C^c}=15$, $\mathcal{L}_{C^v}$ is the MAE loss
calculated between the model output and input label with $\lambda_{C^v}=0.06$,
$\mathcal{L}_{C^m}$ is a cross-entropy loss calculated between the model output
and input label with $\lambda_{C^m}=4$, and $\mathcal{L}_{S}$ is the entropy
regularization on the model $C^m$ output with $\lambda_{S}=2$. See Table
\ref{table:training-results-summary} for a summary of the training results,
demonstrating a good balance between test and training error. A complete log of
training experiments is available at
(\anontext{\url{https://www.comet.ml/ryanhausen/morpheus-deblend/}}).

\begin{table}
  \caption{Training metric results}
  \label{table:training-results-summary}
  \centering
  \begin{tabular}{lll}
    \toprule
    Metric & Training    & Test \\
    \midrule
    MAE              & $27.0183\pm1.0658$ & $28.5090\pm0.3386$ \\
    MSE              & $0.0114\pm0.0001$ & $0.0124\pm0.0006$ \\
    cross-entropy    & $0.9485\pm0.0069$ & $1.0806\pm0.0098$ \\
    \bottomrule
  \end{tabular}
\end{table}

\section{Discussion and Future Work}
\label{sec:discussion}

In this work, we introduced the Partial Attribution Instance Segmentation (PAIS)
scheme for astronomical source debelending. We presented \pcr{} (PCR) as a
method for implementing PAIS within deep learning-based models. We demonstrated
the efficacy of PCR for encoding the results of existing astronomical
deblenders, and developed a novel neural network architecture to recover the PCR
from input flux images. While we demonstrated deblending for single band (F160W)
images, PCR can be extended to multiband images. As with many supervised
methods, our model requires labeled training data. To apply this method on other
survey datasets may require the use of transfer learning
\citep{dominguez_sanchez2019a,pratt1993a} or retraining.

\begin{ack}
\section{Acknowledgements}
\label{sec:acknowledgments}
  RDH would like to thank Roberto Manduchi for helpful conversations.
  BER acknowledges support from NASA contract NNG16PJ25C and grant 80NSSC18K0563.
  The
  authors acknowledge use of the lux supercomputer at UC Santa Cruz, funded by
  NSF MRI grant AST 1828315.
\end{ack}

\section{Broader Impact}
\label{sec:impact-statement}

This work develops a novel method for separating source signals in astronomical
images. Due to the specialized format and problem setting, the authors do not see any
broader negative societal impacts as a result of this work.

\bibliography{refs}

\end{document}